\documentclass[pra,twocolumn,superscriptaddress,10pt,noshowpacs]{revtex4}
\usepackage[english]{babel}
\usepackage[T1]{fontenc}
\usepackage[utf8]{inputenc}
\usepackage{graphicx,epstopdf}
\usepackage{amsmath}

\usepackage{amsfonts}
\usepackage{bbm}
\usepackage{amssymb}
\usepackage{color}
\usepackage{latexsym}
\usepackage{caption}
\usepackage{subcaption}
\usepackage{times,txfonts}

\begin{document}

\title{Black String solutions in Rainbow Gravity}

\author{R. D\'{a}rlla\footnote{E-mail:robertadarlla2@gmail.com}}

\affiliation{${}^{*}$\!Universidade Federal de Campina Grande, Departamento de F\'{i}sica, 63500-000, Campina Grande-PB, Brazil.}

\author{F. A. Brito\footnote{E-mail:fabrito@df.ufcg.edu.br}}

\affiliation{${}^{*}$\!Universidade Federal de Campina Grande, Departamento de F\'{i}sica, 63500-000, Campina Grande-PB, Brazil.}

\author{J. Furtado\footnote{E-mail:job.furtado@ufca.edu.br}}
\affiliation{${}^{*}$\!Universidade Federal do Cariri, Centro de Ci\^{e}ncia e Tecnologia, 63048-080, Juazeiro do Norte-CE, Brazil.}

\date{\today}

\begin{abstract}
    In this paper we study black string solutions under the consideration of rainbow gravity. We have analytically obtained the solution for four-dimensional black strings in terms of the functions $f(E/E_p)$ and $g(E/E_p)$ that sets the energy scale where the rainbow gravity becomes relevant. We have also obtained the Hawking temperature for the black string, from which we could see that the rainbow functions play the role of increasing or decreasing the Hawking temperature for a given horizon radius depending on the choice of such rainbow functions. We have computed the entropy, specific heat and free energy for the black string. The entropy and specific heat exhibit a rainbow dependence, while the free energy is not modified by the rainbow functions. Finally we have studied the effects of the rainbow gravity in the orbits of massive and massless particles around a black string. We could verify that neither massive nor massless particles exhibit stable orbits around a black string in the scenario of rainbow gravity, for any configuration of rainbow functions.    
\end{abstract}

\renewcommand{\thesection}      {\Roman{section}}
\maketitle
\newpage
\section{INTRODUCTION}

There are models of quantum gravity known as Doubly Special Relativity (DSR) that suggests the existence of an invariant energy scale (or lenght) independent of the observer. In such models the dispersion relation is modified when we consider energies near to Planck energy scale \cite{Amelino1,Amelino2,Magueijo,Galan}, which implies that the speed of light is not the only relativistic invariant. Moreover, these models suggest that exists more possible consistent modifications to general relativity other than the introduction of quantum corrections in the Einstein-Hilbert action.

The approaches that receive the name of rainbow's gravity suggest that the usual energy-momentum dispersion relation is deformed close to the Planck scale and that spacetime is also modified due to the non-linear representation of Lorentz transformations, so that its geometry changes according to the energy of the test particle in it. This means that particles with different energies distort spacetime differently in a type of spacetime backreaction leading to the mentioned modification of the relativistic energy-momentum dispersion relation \cite{Camelia2}. These approaches are studied in several scenarios such as string field theory \cite{Samuel}, loop quantum gravity \cite{Pullin}, and non-commutative geometry \cite{Carroll}. Some theoretical proposals suggest corrections both in the action and in the dispersion relation as in \cite{Channuie}.

Some phenomena can be explained through this semi-classical approach, such as the ultra-high-energy cosmic rays that are currently observed but still have unknown origin, suggesting that the dispersion relation is indeed modified. In astrophysics, the influence of the rainbow's gravity on the properties of a black hole has been studied in several scenarios, including its thermodynamics \cite{Leiva,Li,Ali,Valdir1,Feng,Li1,Ronco,Ednaldo,Dehghani,Valdir3,Hamil}, and also in the study of cosmic strings \cite{Valdir1.5,Bakke,Santos}. In addition, to understand the early universe, in which the energies involved were close to the Planck scale, such a modified theory of gravity plays an important role to avoid an initial singularity \cite{Awad,Majumder,Valdir2,Nozari,Hendi,Marc}. Finally, in general field theory there is a lot of recent developments regarding rainbow gravity in the context of Bose-Einstein condensation \cite{Furtado:2021aod}, Klein-Gordon oscillation \cite{Marc}, Landau-Aharonov-Casher effect \cite{Bakke:2022egp}, particle production \cite{Bilim:2023uzf}, among others. 

In this paper we study black string solutions under the consideration of rainbow gravity. We have analytically obtained the solution for four-dimensional black strings in terms of the functions $f(E/E_p)$ and $g(E/E_p)$ that sets the energy scale where the rainbow gravity becomes relevant. We have also obtained the Hawking temperature for the black string, from which we could see that for a give horizon radius the rainbow gravity contribution promotes an increasing in the Hawking temperature.

\section{Rainbow gravity review}

The rainbow gravity was first studied in the context of Double Special Relativity (DSR) and it emerges as a generalization to curved spacetime of the deformed Lorentz symmetry group (locally). One of its consequences is the arising of a modified energy–momentum dispersion relation. Such modification is usually written in the form \cite{Magueijo, Magueijo:2002am, Magueijo:2002xx}
\begin{equation}
    E^2f^2(E/E_P)-p^2c^2g^2(E/E_P)=m^2c^4,
\end{equation}
where $f(E/E_P)$ and $g(E/E_P)$ are the so called rainbow functions, being $E$ the energy of the probe particle and $E_P$ the Planck energy. In the low-energy limit the rainbow functions converges to unit, restoring the standard dispersion relation. However in the high-energy limit the rainbow functions end up violating the usual energy-momentum dispersion relation. The modification of this latter corresponds to a change in the metric, according to \cite{Magueijo:2002xx}, so that the Minkowski spacetime becomes
\begin{equation}
    ds^2=\frac{dt^2}{f^2(E/E_P)}-\frac{1}{g^2(E/E_P)}\delta_{ij}dx^idx^j.
\end{equation}

\begin{figure*}[ht!]
    \centering
\begin{subfigure}{.5\textwidth}
  \centering
  \includegraphics[scale=0.8]{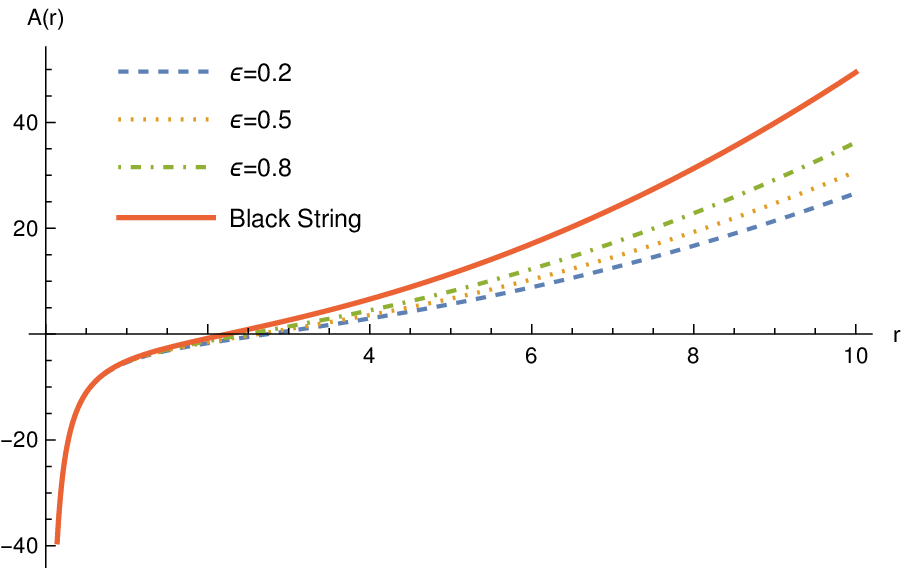}
  \caption{}
 
\end{subfigure}%
\begin{subfigure}{.5\textwidth}
  \centering
  \includegraphics[scale=0.8]{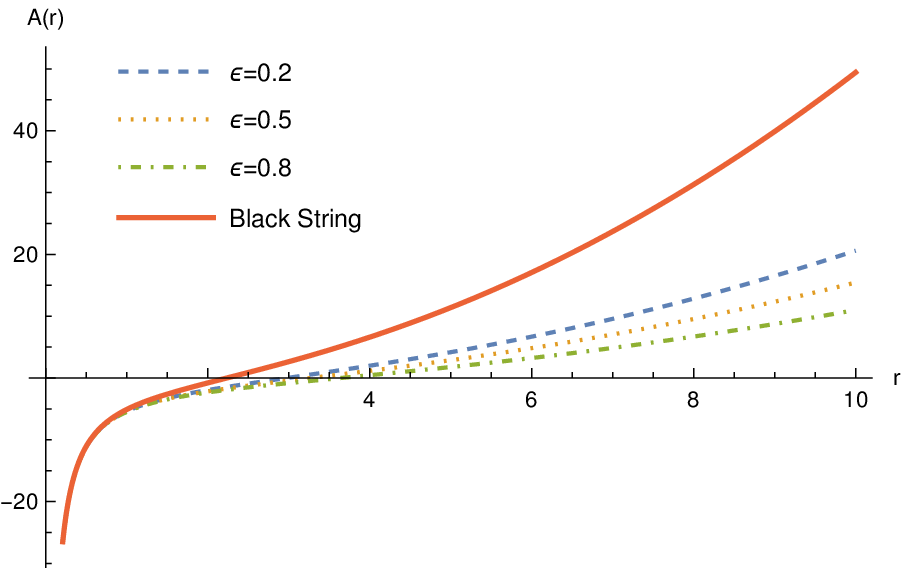}
  \caption{}

\end{subfigure}%
    \caption{Black string solution in rainbow gravity. For this plot we have considered $E_p=1$, $s=1$, $\xi=0.4$, $\alpha=0.5$ and $\mu=0.7$. In (a) we are considering the case I for the rainbow functions while in (b) we are considering the case II.}
    \label{Fig1}
\end{figure*}

In order to study the rainbow gravity effects on the Friedmann-Robertson-Walker (FRW) universe \cite{Awad:2013nxa, Ali:2014xqa}, the following rainbow functions were considered (case I)
\begin{eqnarray}
    f(E/E_P)=1,\,\,\,g(E/E_P)=\sqrt{1-\xi (E/E_P)^s},
\end{eqnarray}
where $s>1$ and $\xi$ is a dimensionless free parameter of the model, which we will consider the same as the other rainbow functions to facilitate comparison between the employed models.

Another interesting choice for the rainbow functions is the following (case II),
\begin{eqnarray}
    f(E/E_P)=g(E/E_P)=\frac{1}{1-\xi(E/E_P)}.
\end{eqnarray}
Such rainbow functions were considered in \cite{Magueijo, Magueijo:2002am} (and references therein) in studying possible nonsingular universe solutions and in \cite{Magueijo:2002xx}, since it assures a constant light velocity, it may provides a solution for the horizon problem.

A last choice of rainbow functions of great interest is given by (case III)
\begin{eqnarray}
    f(E/E_P)=\frac{e^{\xi(E/E_P)}-1}{\xi(E/E_P)},\,\,\, g(E/E_P)=1.
\end{eqnarray}
This choice of the rainbow functions was originally considered in \cite{Camelia2} in the context of Gamma Ray Bursts. Later, this same choice was also addressed in \cite{Awad:2013nxa, Santos:2015sva} in connection with FRW solutions.

\section{Black String solution in Rainbow Gravity}

Let us consider the following line element for the black string 
\begin{eqnarray}\label{23}
    \nonumber ds^2&=&-\frac{A(r)}{f(E/E_p)}dt^2+\frac{1}{g(E/E_p)A(r)}dr^2+\frac{r^2}{g(E/E_p)}d\phi^2\\
    &&+\frac{\alpha^2r^2}{g(E/E_p)}dz^2,
\end{eqnarray}
where $t\in(-\infty, \infty)$, the radial coordinate $r\in[0,\infty)$, the angular coordinate $\phi\in[0,2\pi)$ and the axial coordinate $z\in(-\infty,\infty)$. The $\alpha$ parameter is considered as $\alpha^2=-\Lambda/3$.

\begin{eqnarray}\label{29}
    A(r) = \frac{\alpha^2r^2}{[g(E/E_p)]^2} - \frac{4\mu}{\alpha r} ,
\end{eqnarray}
The above solution for the black string in the rainbow gravity scenario recovers the usual black string solution \cite{Lemos:1994xp} when $g(E/E_p)=1$, i.e., 
\begin{eqnarray}\label{31}
    A(r)=\left(\alpha^2r^2-\frac{4\mu}{\alpha r}\right).
\end{eqnarray}

For the black string the Einstein-Hilbert effective action requires the cosmological constant contribution, so that, 
\begin{eqnarray}\label{24}
    S_u = \frac{1}{2\kappa^2} \int d^4x \sqrt{-g} \left(R-2\Lambda\right).
\end{eqnarray}
where $\kappa=8\pi G$ and $R$ is the Ricci scalar. The EFE for this ansatz gives us
\begin{eqnarray}
    G^t_t -3\alpha^2 &=& [g(E/E_p)]^2\left[\frac{1}{r} \frac{dA(r)}{d r}  + \frac{A(r)}{r^2}\right] - 3\alpha^2, \label{25}
\end{eqnarray} 
We can see that the energy-momentum tensor for the ansatz of equation (\ref{23}) is $T^\mu_\nu = -\rho(r)\text{ diag}(1,1,0,0) + p_l(r)\text{ diag}(0,0,1,1)$, where $p_l = p_\phi = p_z$. This way we can find $A(r)$ by solving $G^t_t -3\alpha^2 = -\kappa^2 \rho(r)$, so that we find

The behaviour of the black string solution in rainbow gravity is depicted in figure (\ref{Fig1}) for the cases I and II. Note that the case III for the rainbow functions does not gives us any modification in the black string solution, since $g(E/E_p)=1$. 

Let us discuss briefly the role played by the rainbow gravity scenario in the black string solution. As we can see in (\ref{Fig1}), as we increase the value of the energy $E$, approaching the Planck energy scale, we also increase the value of the horizon radius for the cases I and II of the rainbow functions.

\section{Black String Thermodynamics in Rainbow Gravity}

\begin{figure*}[ht!]
    \centering
\begin{subfigure}{.5\textwidth}
  \centering
  \includegraphics[scale=0.8]{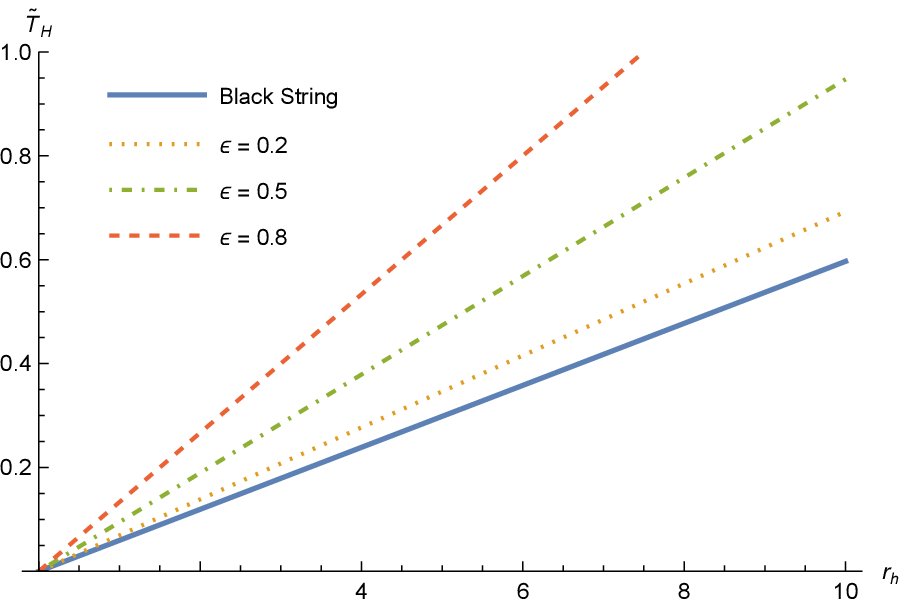}
  \caption{}
 
\end{subfigure}%
\begin{subfigure}{.5\textwidth}
  \centering
  \includegraphics[scale=0.8]{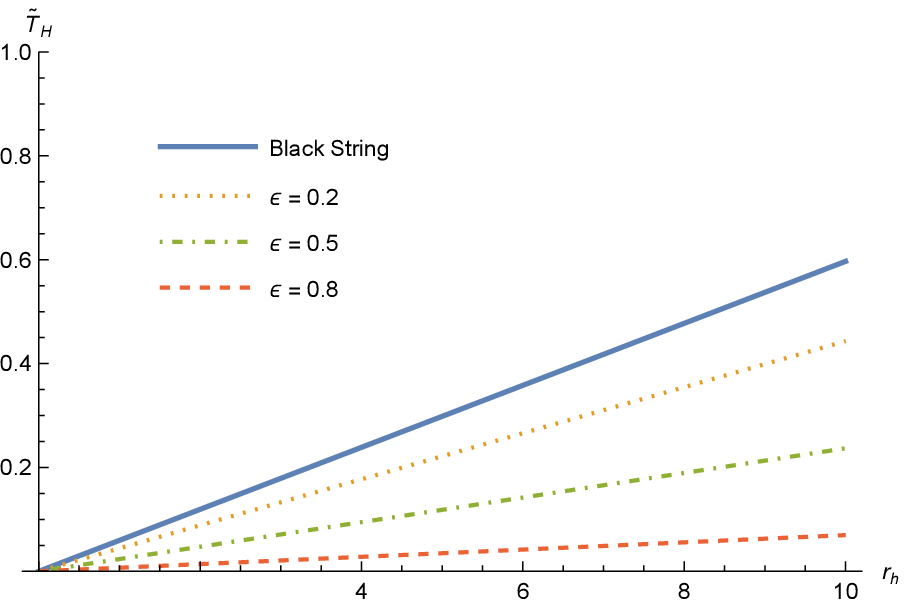}
  \caption{}

\end{subfigure}%
    \caption{Hawking temperature for black string solution in rainbow gravity. For this plot we have considered $E_p=1$, \textcolor{magenta}{$s=1$}, $\xi=0.4$, $\alpha=0.5$ and $\mu=0.7$. In (a) we are considering the case I for the rainbow functions while in (b) we are considering the case II.}
    \label{fig3}
\end{figure*}

Our black string solution in the rainbow gravity scenario has the horizon curves defined by $A(\Tilde{r_h}) = 0$, so that the linear mass can be written as
\begin{eqnarray}
    \mu=\frac{\alpha ^3 \Tilde{r_h}^3}{4 g(E/E_p )^2}.
    \label{linearmass}
\end{eqnarray}
Here $\Tilde{r_h} = r_h \; [g(E/E_P)]^{2/3}$, where $r_{h}$ is the horizon radius of the usual General Relativity solution for the black string. Then, the expression (\ref{linearmass}) becomes
\begin{eqnarray}
    \mu=\frac{\alpha^3 r_{h}^{3}}{4}.
\end{eqnarray}
Thus, this linear mass has no modification due the rainbow gravity.

In possession of the solution for the static black string in the rainbow gravity scenario given by (\ref{29}), we are able to study the thermodynamics of the black string by computing the Hawking’s temperature by means of $T_H=\frac{A'(\Tilde{r}_h)}{4\pi}$. Thus we obtain
\begin{eqnarray}
    \Tilde{T}_H = \frac{3 \alpha^2 \; r_{h}}{4 \pi \; [g(E/E_P)]^{4/3}}.
\end{eqnarray}

\begin{figure*}[ht!]
    \centering
\begin{subfigure}{.5\textwidth}
  \centering
  \includegraphics[scale=0.8]{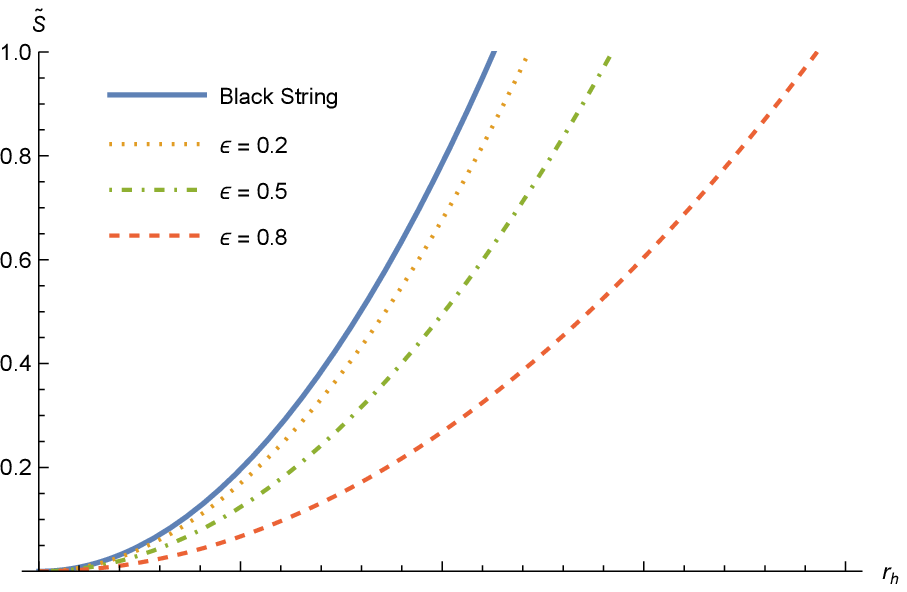}
  \caption{}
 
\end{subfigure}%
\begin{subfigure}{.5\textwidth}
  \centering
  \includegraphics[scale=0.8]{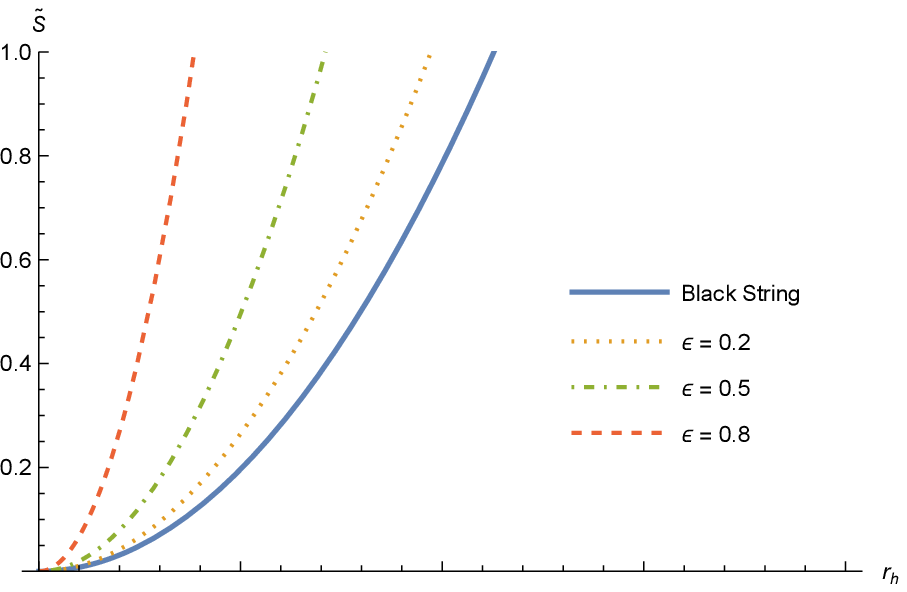}
  \caption{}

\end{subfigure}%
    \caption{Entropy for black string solution in rainbow gravity. For this plot we have considered $E_p=1$, \textcolor{magenta}{$s=1$}, $\xi=0.4$, $\alpha=0.5$ and $\mu=0.7$. In (a) we are considering the case I for the rainbow functions while in (b) we are considering the case II.}
    \label{fig4}
\end{figure*}

The behaviour of the Hawking temperature for the cases I and II is depicted in the figure (\ref{fig3}). For both cases (I and II) the same linear behaviour of the usual Hawking temperature for black strings is present. However some slight differences between the cases must be highlighted. For the case I (fig(\ref{fig3}a)) we can see that for a given horizon radius the Hawking temperature is greater when we consider the effect of rainbow gravity. The opposite occurs for the case II (fig(\ref{fig3}a)), where for a given horizon radius the Hawking temperature is smaller when we consider the effect of rainbow gravity.

In order to properly understand the thermodynamics of the black string in the rainbow gravity context it is necessary to compute the entropy, specific heat and free energy. The entropy can computed directly from the expression $dS=\frac{d\mu}{\Tilde{T}_H}$, in which we get
\begin{equation}
\Tilde{S} = \frac{\pi \; \alpha \; r_{h}^{2} \; [g(E/E_P)]^{4/3}}{2}.
\end{equation}
Cleary, this recovers the usual black string result $S=\frac{1}{2} \pi  \alpha  r_h^2$ when $g(E/E_P) = 1$. As we can see in figure (\ref{fig4}), for both cases we have the same quadratic dependence of the horizon radius that the usual black string entropy exhibit. However, differently from the Hawking temperature, the case I promotes a decreasing in the entropy for a given horizon radius while the case II promotes an increasing in the entropy for a given horizon radius.

The specific heat can be calculated by $\Tilde{C}_v=\frac{d\mu}{d\Tilde{T_H}}$ from which we obtain
 \begin{equation}
\Tilde{C}_v = \pi \alpha r_{h}^{2} \; [g(E/E_P)]^{4/3}
 \end{equation}

Similar to entropy, in case I for a given horizon radius the specific heat is smaller when we consider the effect of rainbow gravity. The opposite happens for case II. When $g(E/E_P) = 1$ we get $C_v=\pi  \alpha  r_h^2$, i.e. the usual black string specific heat in General Relativity. The behaviour of the specific heat for the cases I and II of
the rainbow functions is depicted in (\ref{fig5}). As it is widely known, the thermodynamical stability of black holes (black strings for our case) is directly related to the sign of the heat capacity. A positive heat capacity indicates that the system is thermodynamically stable, while its negativity imply a thermodynamical instability. Therefore, the result for the specific heat in the context of rainbow gravity indicates a thermodynamically stable black string. 

\begin{figure*}[ht!]
    \centering
\begin{subfigure}{.5\textwidth}
  \centering
  \includegraphics[scale=0.8]{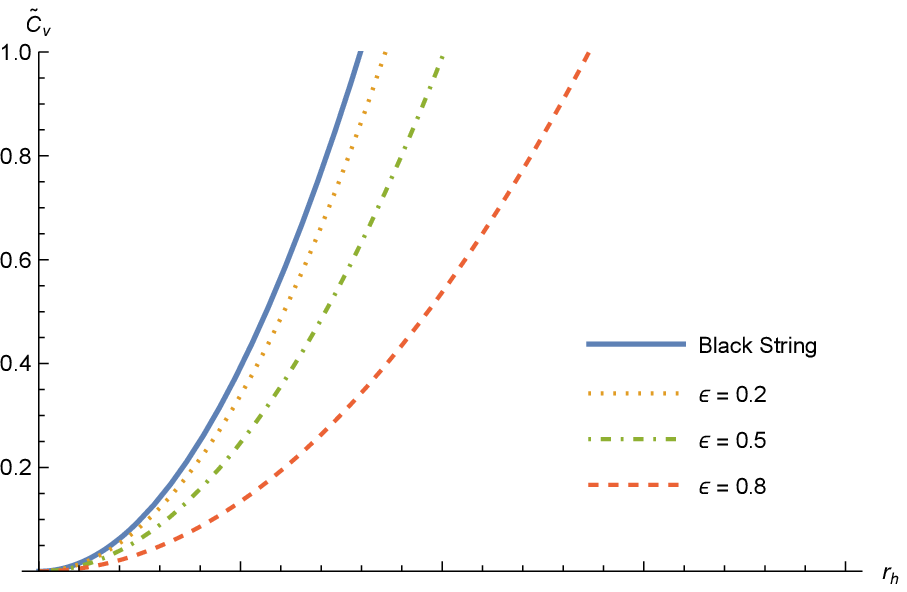}
  \caption{}
 
\end{subfigure}%
\begin{subfigure}{.5\textwidth}
  \centering
  \includegraphics[scale=0.8]{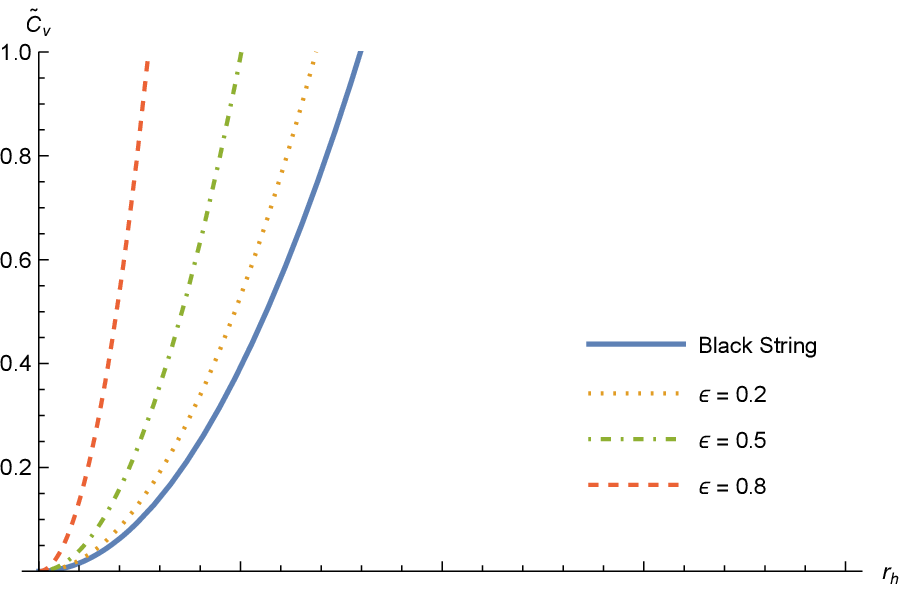}
  \caption{}

\end{subfigure}%
    \caption{Especific Heat for black string solution in rainbow gravity. For this plot we have considered $E_p=1$, \textcolor{magenta}{$s=1$}, $\xi=0.4$, $\alpha=0.5$ and $\mu=0.7$. In (a) we are considering the case I for the rainbow functions while in (b) we are considering the case II.}
    \label{fig5}
\end{figure*}

On the other hand, the rainbow gravity presents no modification in the free energy $F=\mu-T_H S$, yielding therefore the usual black string
result
\begin{eqnarray}
    F = -\frac{\alpha ^3 r_h^3}{8}.
\end{eqnarray}

\section{Geodesics and circular orbits}
The particle's geodesic in orbit around a static black string is given by
\begin{eqnarray}
    \dot{r}^2=\omega^2-A(r)\left(\frac{L^2}{r^2}+m^2\right),
\end{eqnarray}
where $\omega$ is the particle's energy, $L$ is the angular momentum and $m$ is the particle's mass. Thus the effective potential is defined as
\begin{eqnarray}
    V_r=A(r)\left(\frac{L^2}{r^2}+m^2\right).
\end{eqnarray}

The circular geodesics occur at the points $r_c$ satisfying $\frac{1}{2}\dot{r}_c^2=0$ and $V_r'(r_c)=0$. In Fig. (\ref{fig5}) we depict the effective potential of massless and massive particles for the black string in the rainbow gravity scenario. It is shown that there is no case where circular orbits are stable, similarly to the usual black string solution. Therefore, the rainbow gravity does not modify significantly the results for geodesics and circular orbits in comparison to the usual black string.

\begin{figure*}[ht!]
    \centering
\begin{subfigure}{.5\textwidth}
  \centering
  \includegraphics[scale=0.8]{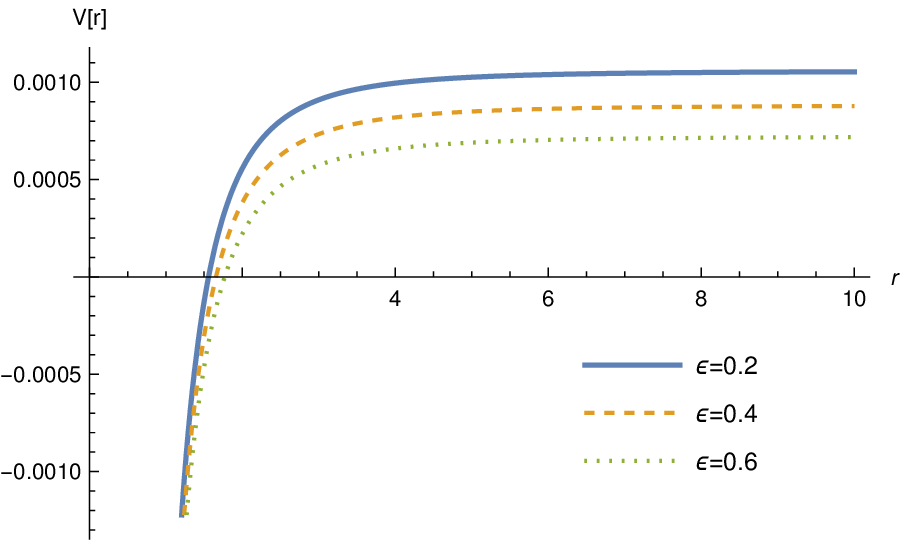}
  \caption{}
 
\end{subfigure}%
\begin{subfigure}{.5\textwidth}
  \centering
  \includegraphics[scale=0.8]{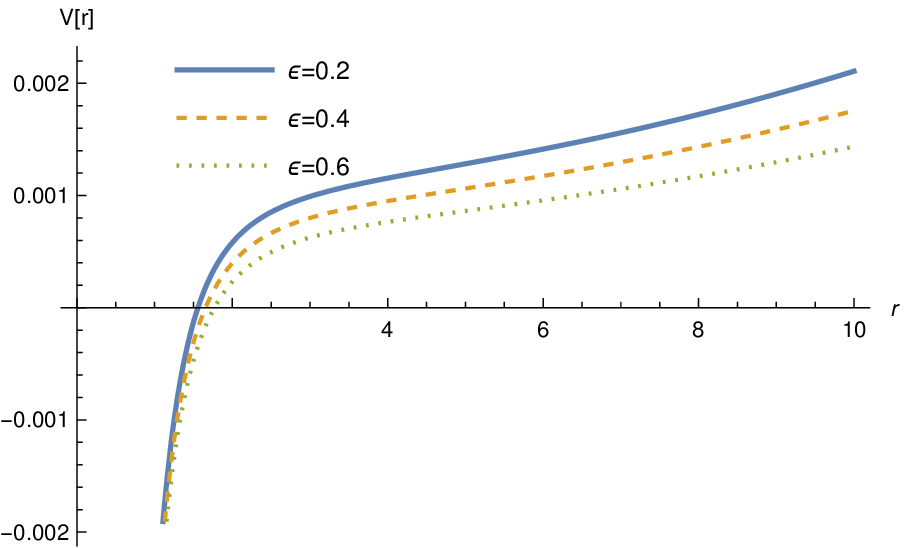}
  \caption{}

\end{subfigure}%
    \caption{Effective potential (a) for massless particles and (b) for massive particles. For this plot we have considered $E_p=1$, $s=1$, $\xi=0.4$, $\alpha=0.5$, $\mu=0.7$ and $L=0.1$.}
    \label{fig5}
\end{figure*}

\section{Conclusion}

In this paper we study black string solutions under the consideration of rainbow gravity. We have analytically obtained the solution for four-dimensional black strings in terms of the functions $f(E/E_p)$ and $g(E/E_p)$ that sets the energy scale where the rainbow gravity becomes relevant. We could verify that the black string solution depends only on the function $g(E/E_p)$, and consequently, all the thermodynamic parameters will depend only on $g(E/E_p)$. We have plotted the behaviour of the black string solution in (\ref{Fig1}), and we could see that as we increase the value of the energy $E$, approaching the Planck energy scale, we also increase the value of the horizon radius for the cases I and II of the rainbow functions.

We have also obtained the Hawking temperature for the black string, from which we could see that the rainbow functions play the role of increasing or decreasing the Hawking temperature for a given horizon radius depending on the choice of such rainbow functions. We have computed the entropy, specific heat and free energy for the black string. The entropy and specific heat exhibit a rainbow dependence, while the free energy is not modified by the rainbow functions.

Finally we have studied the effects of the rainbow gravity in the orbits of massive and massless particles around a black string. We could verify that neither massive nor massless particles exhibit stable orbits around a black string in the scenario of rainbow gravity, for any configuration of rainbow functions. 

\section*{Acknowledgements}

FAB would like to thank CNPq and CNPq/PRONEX/FAPESQ-PB (Grant nos. 165/2018 and 312104/2018-9), for partial financial support. JF would like to thank the Fundação Cearense de Apoio ao Desenvolvimento Cient\'{i}fico e Tecnol\'{o}gico (FUNCAP) under the grant PRONEM PNE0112-00085.01.00/16 for financial support.



\end{document}